\documentclass[12pt,preprint]{aastex}

\newcommand{\myemail}{quanz@mpia.mpg-hd.de}

\slugcomment{To appear in ApJ}

\shorttitle{emph{Spitzer} observations of RNO 1B/1C}
\shortauthors{Quanz et al.}

\begin{document}


\title{Deeply embedded objects and shocked molecular hydrogen: The environment of the FU Orionis stars RNO 1B/1C}


\author{S. P. Quanz, Th. Henning, J. Bouwman, H. Linz}
\affil{Max Planck Institute for Astronomy, K\"onigstuhl 17, 69117 Heidelberg,
    Germany}
\email{\myemail}
\and
\author{F. Lahuis}
\affil{SRON Netherlands Institute for Space Research, P.O. Box 800, 9700 AV Groningen, Netherlands}






\begin{abstract}
We present \emph{Spitzer IRAC} and \emph{IRS} observations of the dark cloud L1287. 
The mid--infrared (MIR) \emph{IRAC} images show 
deeply embedded infrared sources in the vicinity of the FU Orionis objects RNO 1B and RNO 1C 
suggesting their association with a small young stellar cluster. For the first time we resolve 
the MIR point source associated with IRAS 00338+6312 
which is a deeply embedded intermediate--mass protostar driving
a known molecular outflow. The \emph{IRAC} colors of all objects are consistent 
with young stars ranging from deeply embedded Class 0/I sources to Class II objects, part of which
appear to be locally reddened. 
The two \emph{IRS} spectra show strong absorption bands by ices and dust particles, 
confirming that the circumstellar environment around RNO 1B/1C has a high optical depth.
Additional hydrogen emission lines from pure rotational transitions are superimposed on the spectra.
Given the outflow direction, we attribute these emission lines
to shocked gas in the molecular outflow powered by IRAS 00338+6312. 
The derived shock temperatures are in agreement with high velocity C-type shocks. 

\end{abstract}





\section{Introduction}
The dark cloud L1287 contains the galactic nebulae GN 00.34.0 (associated with the young F-type star 
RNO 1) and GN 00.33.9 which harbors the IRAS point source IRAS 00338+6312 \citep[$d=800$ pc, ][]{persi1988}. 
Two young stars, RNO 1B and RNO 1C\footnote{We use the same nomenclature as \citet{staudeneckel1991}.}, lie 
slightly south-west of the catalog position of the 
IRAS source. Those objects show
properties of FU Orionis objects (FUORs): (1) \citet{staudeneckel1991} found that RNO 1B 
brightened by at least 3 mag over a period of 12 years and that it showed a variable and blueshifted
P Cygni profile in H$\alpha$ and additional broad and double--peaked absorption lines in its optical spectrum;
(2) \citet{kenyon1993} obtained near--infrared spectroscopic data for both objects and found the FUOR--typical strong
2.3$\,\mu$m CO absorption bands.
As the error ellipse of the IRAS source includes both FUORs it was long debated whether 
there is still another deeply embedded source close to the FUOR objects or whether IRAS just 
measured the integrated flux from these two objects. From high--resolution near--infrared (NIR) 
polarimetric maps \citet{weintraubkastner1993} concluded that an additional embedded object should
be present close to the location of the IRAS source. This idea was supported by the discovery of a 
3.6 cm continuum peak \citep{anglada1994} and an H$_2$O maser \citep{fiebig1995}, both nearly coinciding
with the IRAS position.
A bipolar outflow in the region was found by \citet{snell1990} and later confirmed by \citet{yang1991}. 
As the positions of the IRAS source and the FUORs line up along the outflow axis it has long
been uncertain which source was driving the outflow. From interferometric observations in CS \citet{yang1995} suggested that the IRAS source was most likely the driving source. 
Recently, \citet{xu2006} mapped the outflow in CO and came to the same conclusion. However, \citet{mcmuldroch1995} presented (sub--)millimeter observations favoring RNO 1C 
as the outflow driving source.

In this paper we present mid--infrared (MIR) imaging and 
spectroscopy data taken with the \emph{IRAC} and \emph{IRS} instruments 
onboard the \emph{Spitzer Space Telescope}. For the first time we resolve the MIR point source 
associated with the IRAS source and find additional, partly deeply embedded, objects.
The two \emph{IRS} spectra probe the composition of the dense circumstellar environment in the 
vicinity of RNO 1B/1C and bear additional traces of shocked H$_2$ gas.



\section{Observations and Data Reduction}
\begin{figure}
 \vspace{1.5cm}
\centering
   \includegraphics[scale=.5]{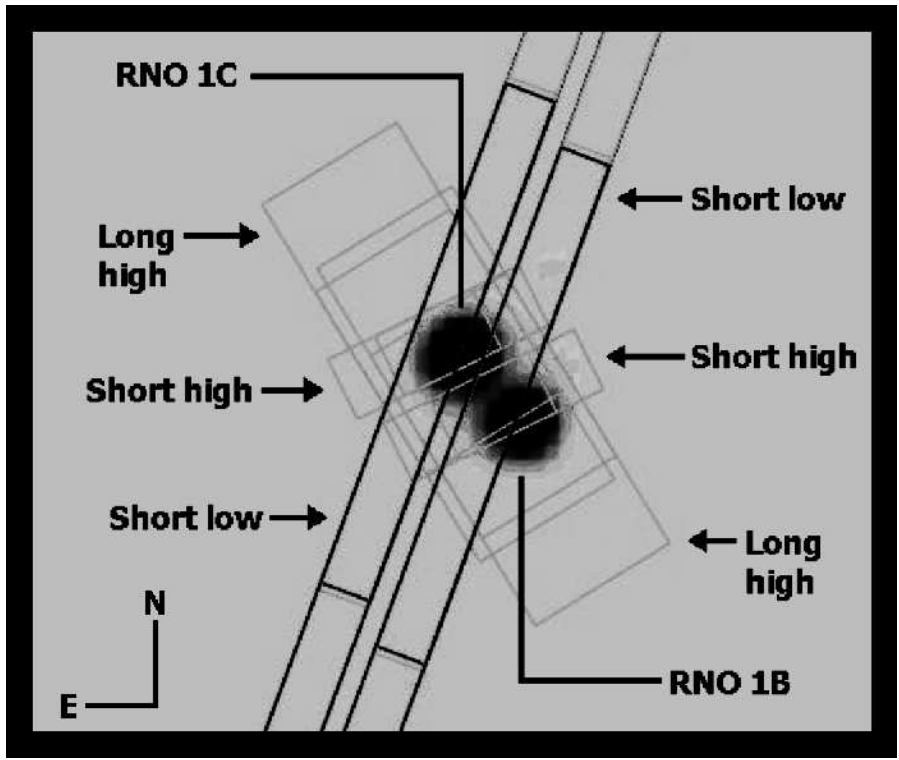}
   \caption{The slits of the different modules of the \emph{Spitzer/IRS} overplotted on the inverted 2MASS Ks-filter
   image. The two FUOR objects RNO 1B and RNO 1C are not centered in the slits (north is up, east to the left).}
 \label{slits}%
\end{figure}

All data were part of the GTO program by R. Gehrz and 
publicly available from the \emph{Spitzer} data archive. 
An overview of the observational setup and the datasets is provided in Table~\ref{journal}. 
The \emph{IRAC} images were obtained in sub--array mode leading to an effective field--of--view
of ~40$''$ centered on the position given in Table~\ref{journal}. 
The spectra taken with \emph{IRS} cover the wavelength range $5-37$ $\,\mu$m.
Overplotting the spectral slits of \emph{IRS}
on the 2MASS Ks-filter image reveals that apparently RNO 1B and RNO 1C were not centered within 
the slits (Fig.~\ref{slits}). 
The short low--resolution spectrum ($5.2-14.5$ $\,\mu$m, R $\sim 64-128$) and the short 
high--resolution spectrum ($9.9-19.6$ $\,\mu$m, R $\sim 600$) close to RNO 1B seems to probe mainly flux coming
from in between the two objects. The spectrum close to RNO 1C contains presumably also less flux 
than expected due to the slight mispointing. The long high--resolution part in either spectrum 
($18.7-37.2$ $\,\mu$m, R $\sim 600$) includes flux from both components 
and also additional flux from the IRAS source as all three objects lie within the slit of the spectrograph.

The \emph{IRAC} images were reduced with the MOPEX package provided by the \emph{Spitzer Science Center}. Interpolation, outlier detection and 
co--addition of the images were carried out for each filter individually. The astrometry was refined by comparing the position of detected 
sources to known 2MASS objects. The photometry was carried out with the {\tt daophot} package provided within the {\tt IRAF\footnote{http://iraf.noao.edu/}} environment. 
As there was only a limited number of sources available within the small field covered by the camera 
we measured the point-spread function (PSF) of 
RNO 1C and used it as reference to do PSF photometry for all the other sources. Following the \emph{IRAC} Data Handbook 5.1.1 we converted the
pixel values from MJy\,sr$^{-1}$ to DN\,s$^{-1}$ and computed the corresponding magnitudes via $m =-2.5\,log(x)+\Delta_{ZP}$ with $x$ denoting the flux measured in DN\,s$^{-1}$ and $\Delta_{ZP}$ being the zero point for each filter\footnote{Zero points taken from \citet{hartmann2005}: 19.66 (3.6$\,\mu m$), 18.94 (4.5$\,\mu m$), 16.88 (5.8$\,\mu m$), 17.39 (8$\,\mu m$).}. 
For the initial PSF fit we used a PSF size of 2 pixel and applied aperture corrections as described in the Data Handbook to obtain the final magnitudes. 

Our final \emph{IRS} spectra are based on the intermediate 
{\tt droopres} (for the low--resolution data) and {\tt rsc} (for the high--resolution data) 
products processed through the S13.2.0 
version of the {\it Spitzer\,} data pipeline. Partially derived from 
the {\tt SMART} software package \citep{higdon2004} these intermediate 
data products were further processed using spectral extraction tools 
developed by the "Formation and Evolution of Planetary Systems" 
(FEPS) {\it Spitzer\,} science legacy team (see, Explanatory Supplement v. 3.0 within 
the FEPS data deliveries\footnote{http://ssc.spitzer.caltech.edu/legacy/fepshistory.html}. 

For the short low--resolution observations, the spectra where extracted using a 6.0 pixel
fixed-width aperture in the spatial dimension resulting in a 39.96 arcsec$^2$ extraction aperture
on the sky. The background was subtracted 
using associated pairs of imaged spectra from the two nodded positions 
along the slit. This process also subtracts stray light contamination 
from the peak-up apertures and adjusts pixels with anomalous dark 
current relative to the reference dark frames. Pixels flagged by 
the {\it Spitzer\,} data pipeline as being "bad" were replaced with 
a value interpolated from an 8 pixel perimeter surrounding the errant 
pixel. The high--resolution spectra were extracted with the full aperture size (53.11 arcsec$^2$ for the
short--high and 247.53 arcsec$^2$ for the long--high module). The sky 
contribution was estimated by fitting a continuum to the data. 
Both, the low-- and high--resolution spectra, are calibrated using a spectral response function derived from 
IRS spectra appropriate for the two nod positions in the slit at which we extract the specta, 
and Cohen or MARCS stellar models for a suite of calibrators provided by the {\it Spitzer\,Science\,Centre}. 
The multiple orders in our spectra match to within 10\%. These small flux offsets are likely related 
to small pointing offsets. We computed correction factors for possible flux loss due 
to telescope mispointing based on the PSF of the \emph{IRS} instrument. However, accurate results can only
be obtained if the flux is dominated by one source.
The offsets also result in
low level fringing at wavlengths longer than 20 micron in the low--resolution spectra and at all wavelengths
in the high--resolution spectra. We removed these fringes using the {\tt irsfringe} package developed by
F. Lahuis. The relative errors between spectral points within one order are dominated by the 
noise on each individual point and not by the calibration. Based on our experience with \emph{Spitzer IRS} data
we estimate a relative flux calibration error across a spectral order of 
$\approx 5$~\% and an absolute calibration error between 
orders/modules of $\approx 10$~\%. These values, however, are based on point source observations with 
accurate telescope pointing. The data presented here are more complicated as at least part of the 
emission is moderately extended,
multiple sources contribute to the observed fluxes, each contribution is wavelength dependent, and the
telescope slits were not centered on the main objects. Thus, an accurate flux calibration between the different   
modules is difficult and apparent offsets are further discussed in section 3.2.






\section{Results}
\subsection{IRAC photometry}
\begin{figure}
 \vspace{1.5cm}
\centering
   \includegraphics[width=17cm,angle=0]{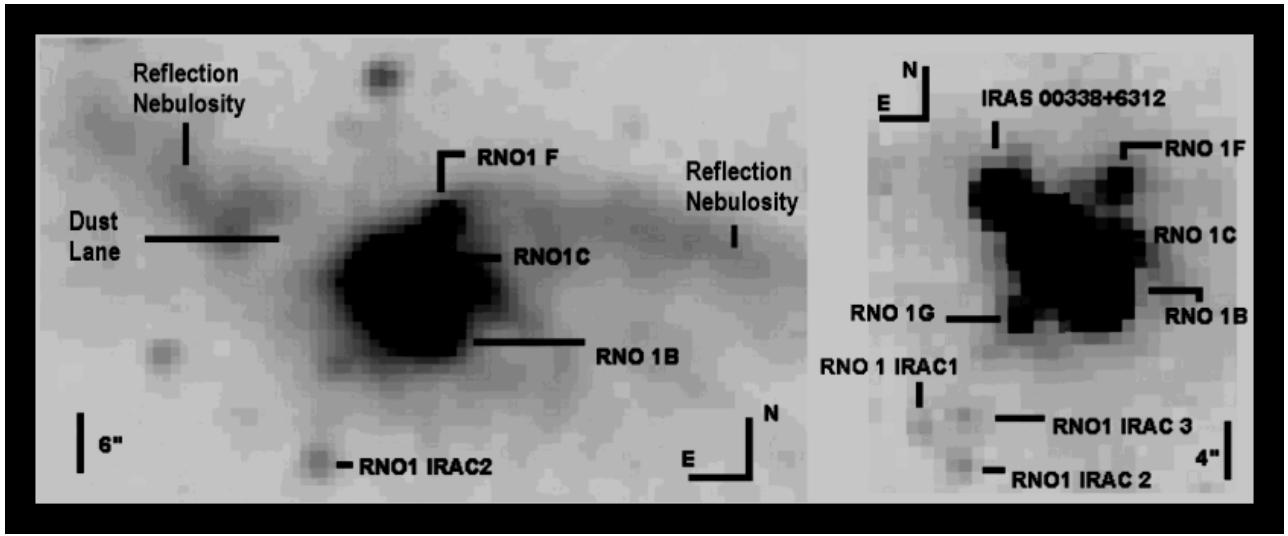}
   \caption{The RNO 1B/1C region in the 2MASS Ks-filter (left) and the 5.8$\,\mu$m \emph{IRAC} filter (right).
   The images are inverted and the contrast is set to make also fainter objects visible. 
   The \emph{IRAC} image contains more sources and  
   especially the IRAS source clearly shows up at 5.8$\,\mu$m while it seems hidden behind a dust lane in 
   the 2MASS image. The magnitudes of the individual sources are listed in Table~\ref{irac_fluxes}. }
 \label{fig1}%
\end{figure}

Within the limited field--of--view of the \emph{IRAC} subarray mode we identified 8 sources
that were detected in at least three of the four \emph{IRAC} bands.  
Figure~\ref{fig1} provides a comparison between the RNO 1B/1C region as seen in the 2MASS Ks-filter and the
5.8$\,\mu$m \emph{IRAC} filter. For the first time the MIR point source related to
IRAS 00338+6312 is detected in the \emph{IRAC} band. Additional fainter objects are also 
present, some of which are detected for the
first time. Fig.~\ref{fig9} shows a color composite image of the
RNO 1B/1C complex based on three \emph{IRAC} filters. 
Table~\ref{irac_fluxes} lists all objects that were 
identified in at least three \emph{IRAC} bands and summarizes the derived fluxes. 
The errors are based on the results from the PSF-photometry. Since we used the PSF of RNO 1C as reference PSF
the corresponding errors are relatively small compared to the other objects.  

\begin{figure}
 \vspace{1.5cm}
\centering
    \includegraphics[scale=0.5,angle=90.]{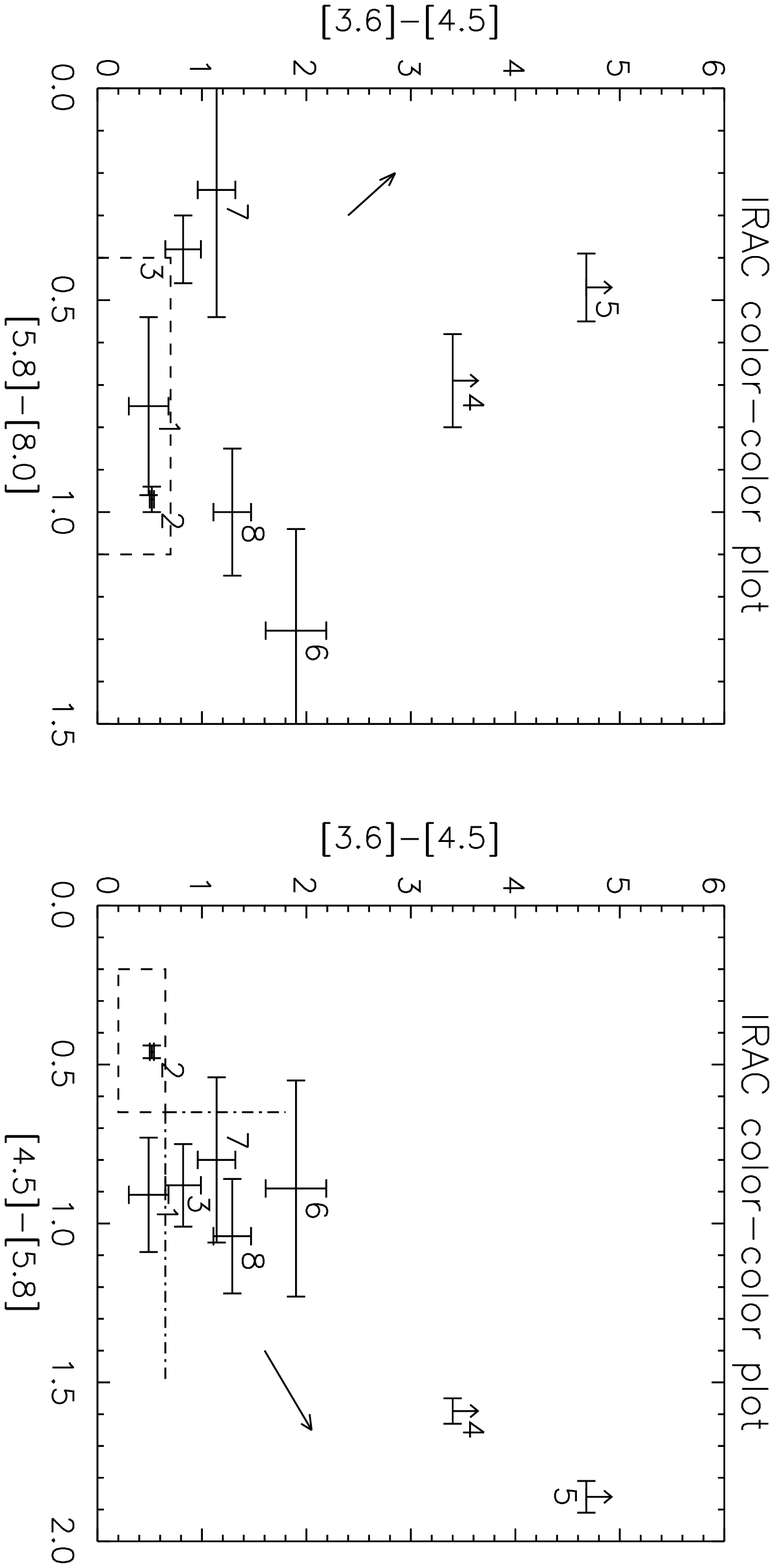}
   \caption{\emph{IRAC} color-color plots for the objects listed in Table~\ref{irac_fluxes}. 
   The dashed boxes indicate the regions corresponding to Class II objects in \citet{hartmann2005}. The 
   dashed-dotted region in the right plot defines the position of Class 0/I sources in \citet{hartmann2005}.
   In the left plot the Class 0/I objects from \citet{hartmann2005} have the same [5.8]-[8.0] colors as 
   the Class II objects but redder [3.6]-[4.5] colors. Thus, they lie "above" the dashed box.
   The reddening vectors correspond to $A_V=30$ mag and are based on a Vega-like 
   spectrum and the reddening law given by \citet{mathis1990}. }. 
 \label{fig2}%
\end{figure}

IRAS 00338+6312 and RNO 1G\footnote{The object we call RNO 1G appears to be identical to 
the embedded YSO from \citet{weintraubkastner1993}.}
were not detected in the shortest \emph{IRAC} band at 3.6$\,\mu$m. 
The 2MASS image in Fig.~\ref{fig1} shows that at least 
the IRAS source is deeply embedded in a dense dusty environment explaining the non-detection at this wavelength. 
Fig.~\ref{fig2} shows two color--color plots based on the
four \emph{IRAC} bands for all objects listed in Table~\ref{irac_fluxes}. 
Following \cite{hartmann2005} who analyzed a large sample of pre-main sequence stars in
the Taurus star-forming region we use these plots to
classify the different objects. The colors of the reddest objects (IRAS 00338+6312 and RNO 1G) are consistent 
with very young protostars. The colors of the newly discovered objects RNO1 IRAC1 and RNO1 IRAC3 are consistent with 
Class 0/I systems although it has to be confirmed that these are indeed nearby objects and not highly reddened background 
sources. While the colors of the FUOR object RNO 1C fit in the region of Class II objects from \cite{hartmann2005}
in both plots (Fig.~\ref{fig2}), 
RNO 1B fits only in one color-color diagram into this regime. In the right plot of Fig.~\ref{fig2}
RNO 1B is too red for a Class II source in the [4.5]-[5.8] color, but not red enough in the
[3.6]-[4.5] color to be a Class 0/I object. 
Finally, RNO1 IRAC 2 and RNO 1F, which both show up in the 2MASS Ks-band image, can also be interpreted as Class II objects. However, the colors of RNO1 IRAC2, but partly also of RNO 1F, 
are possibly altered due to local extinction effects. 

\subsection{IRS spectroscopy}
\begin{figure}
 \vspace{1.5cm}
\centering
  \epsscale{1.0}
 \plotone{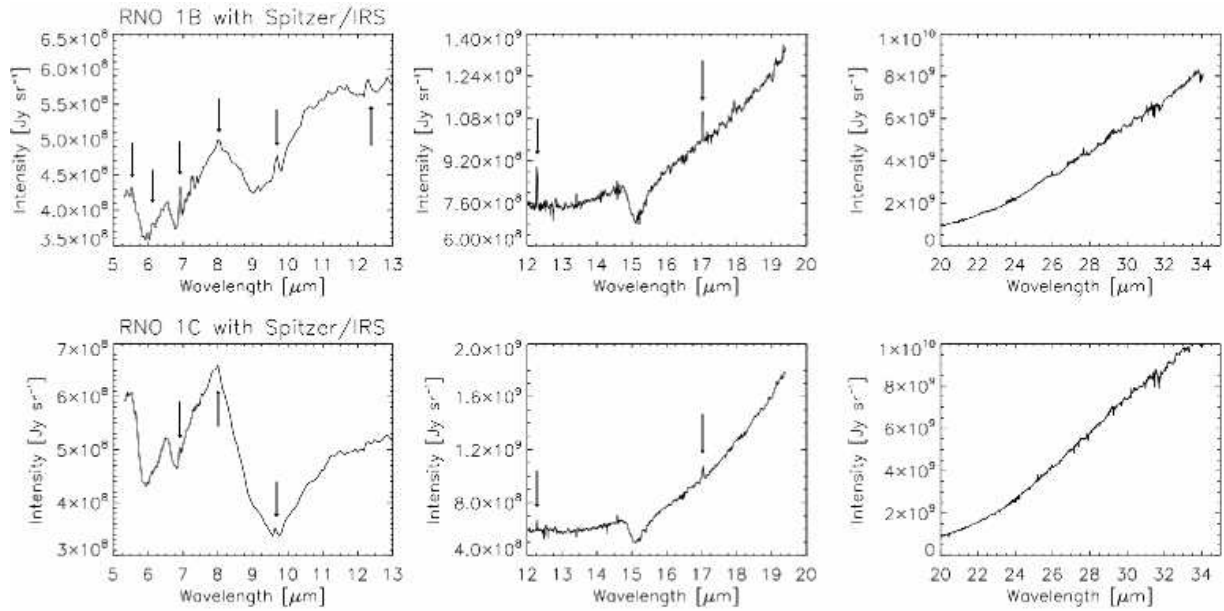}
   \caption{\emph{Spitzer/IRS} spectra close to RNO 1B (upper row) and RNO 1C (lower row). 
   The leftmost column shows the low--resolution data while the other 
   two columns present the high--resolution spectra.
   The positions of the H$_2$-emission lines are indicated by the arrows.}. 
 \label{fig3}%
\end{figure}

In Figure~\ref{fig3} we show the \emph{Spitzer IRS} spectra without any extinction correction.
As pointed out earlier the spectral slits of the short wavelength modules were apparently not 
directly centered on the objects RNO 1B and RNO 1C and thus the measured 
fluxes cannot be attributed to these objects with very high accuracy. 
Also, there seems to be an offset in the flux in 
the high--resolution regime compared to the low--resolution 
part of the spectrum. 
To see to which extent this effect is caused by different aperture sizes we plotted the spectra in 
intensities (i.e., flux density per solid angle) rather then in Jansky per wavelength. 
It shows, however, that even after this correction significant offsets remain: 
Around 13$\,\mu$m the difference between the low--resolution spectrum and the short high--resolution spectrum 
amounts to a factor of $\approx$$\,1.25$ and $\approx$$\,1.08$ for RNO 1B and RNO 1C, respectively, with 
the high--resolution part showing higher intensities. At least for the RNO1 B spectrum this offset is larger
than normally expected from the calibration accuracy. We thus believe that part of this offset can be attributed to
the different orientations of the spectral slits on the sky probing different regions of the extended emission.
The offset between the short and long wavelength range of the high--resolution data corresponds to factors of
$\approx$$\,1.5$ and $\approx$$\,2.0$ at 20$\,\mu$m for RNO 1B and RNO 1C, respectively. Also here the short high--resolution 
spectrum shows higher intensities. This can be explained by the approximately five times smaller aperture in the short high--resolution 
module. Although this aperture is probing a significantly smaller regions on the sky these regions do presumably contribute in total to flux 
in the above mentioned wavelength regime. The large aperture of the long high--resolution module on the other hand is certainly also
probing regions without any significant flux (see also, Fig.~\ref{slits}).

\subsubsection{Ices and silicates}
\begin{figure}
 \vspace{1.5cm}
\centering
  \epsscale{0.9}
 \plotone{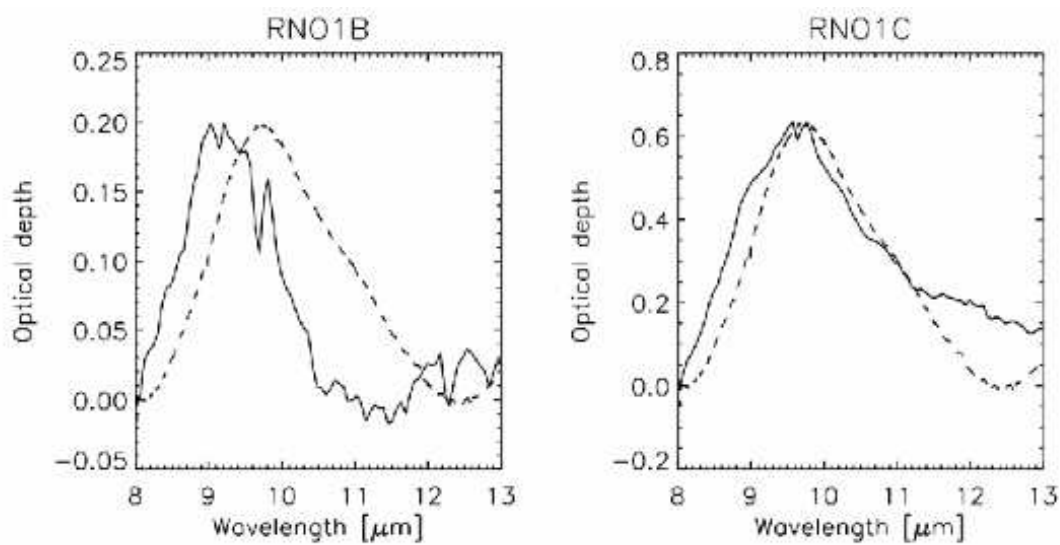}
   \caption{Comparison of the optical depth of the 10$\,\mu$m absorption feature 
   with a typical ISM absorption feature. The solid lines show the optical depth observed
   towards the RNO complex while the dashed lines represent the (scaled) optical depth towards the
   galactic center \citep{kemper2004}. The absorption in the RNO region can not be solely caused by pure 
   ISM silicate grains as both absorption features seem to be slightly shifted towards shorter wavelengths.
   Furthermore, while the spectrum close to RNO 1C shows additional absorption longwards of 11$\,\mu$m (possibly
   blended with H$_2$O ice at 13$\,\mu$m), the feature close to RNO 1B is too narrow and an additional
   emission feature might contribute longwards of $\sim$10.5$\,\mu$m.} 
 \label{10mu_absorption}%
\end{figure}

At 6$\,\mu$m both spectra show clear water ice absorption bands. At slightly longer
wavelength ($\sim$6.85$\,\mu$m) additional absorption 
possibly arising from CH$_3$OH ice is present \citep{dishoeck2004}. 
The 10$\,\mu$m silicate band is also seen in absorption in both spectra (stronger close to RNO 1C), although the shape of the feature differs from the absorption feature 
caused by typical interstellar medium silicate grains. To analyze the differences in more detail we
fitted a continuum to the 10$\,\mu$m region of the spectra and computed the optical depth (Fig~\ref{10mu_absorption}). 
In both cases the peak of the absorption is slightly shifted towards shorter 
wavelengths indicating a non-ISM like dust composition. Additional absorption (RNO 1C) or 
possibly additional dust emission on top of the absorption feature (RNO 1B) is seen at longer
wavelengths. At 15.2$\,\mu$m CO$_2$ ice is creating a prominent absorption band and at $\sim$18$\,\mu$m 
additional silicate absorption seems to be present.
All these features
indicate the existence of a dense dusty and icy environment in which the two FUORs are embedded. A further analysis of these features also for a larger
sample of FUORs will be presented in an upcoming paper (Quanz et al. in preparation).

\subsubsection{Pure rotational H$_2$ emission}
In addition to the ice and silicate features, H$_2$ emission lines from purely rotational 
quadrupole transitions are present in both spectra (Fig.~\ref{fig3}). 
While in the spectrum close to RNO 1B all transitions from S(1) to S(7) can be identified, 
the spectrum close to RNO 1C shows only the lines from S(1) to S(5). This can be explained
with the higher continuum flux close to the S(6) and S(7) line in the latter spectrum and with 
apparently lower excitation temperatures (see below). 
The lowest transition S(0) near 28.22$\,\mu$m is not detected in either spectrum. This is due to the 
strongly rising continuum at longer wavelengths. Here, the spectral slit contained flux from both RNO objects and also from the deeply embedded IRAS source so that the continuum emission and the related flux errors completely dominate a possible weak emission line. A detailed analysis is provided in the appendix.  

\begin{figure}
 \vspace{1.5cm}
\centering
  \epsscale{1.0}
 \plotone{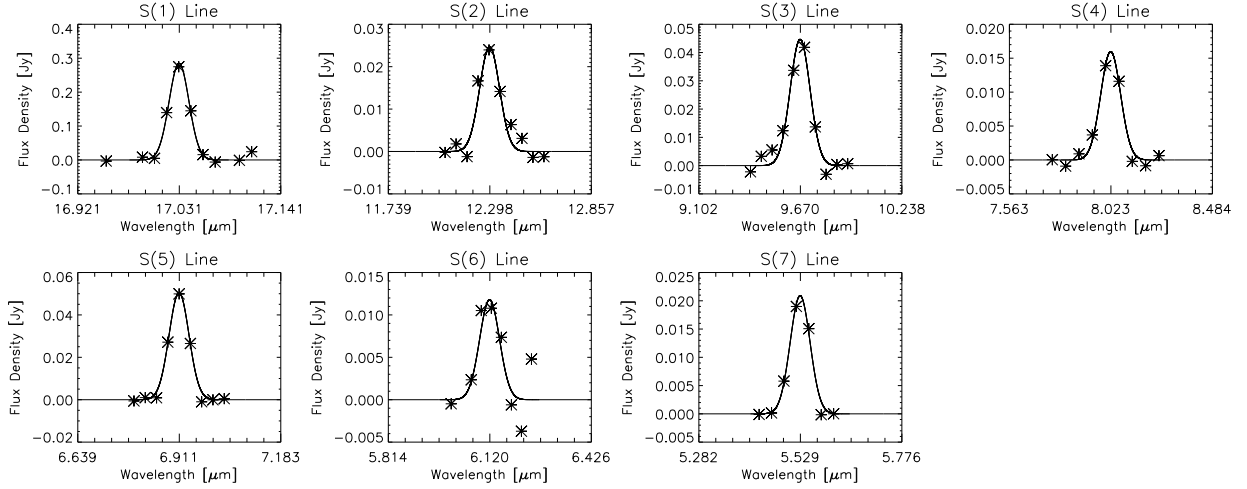}
   \caption{Gaussian fits to the H$_2$ lines in the spectrum close to RNO 1B after continuum 
   subtraction. The line at 12.28$\,\mu$m shown here was measured in the low--resolution spectrum. 
   Table~\ref{hydrogen_lines}, however, contains the flux of this line  
   measured in both, the low-- and high--resolution part of the spectrum. The line at 6.11$\,\mu$m
   is blended with the ice feature in this region. }
 \label{fig4}%
\end{figure}

\begin{figure}
 \vspace{1.5cm}
\centering
  \epsscale{1.0}
 \plotone{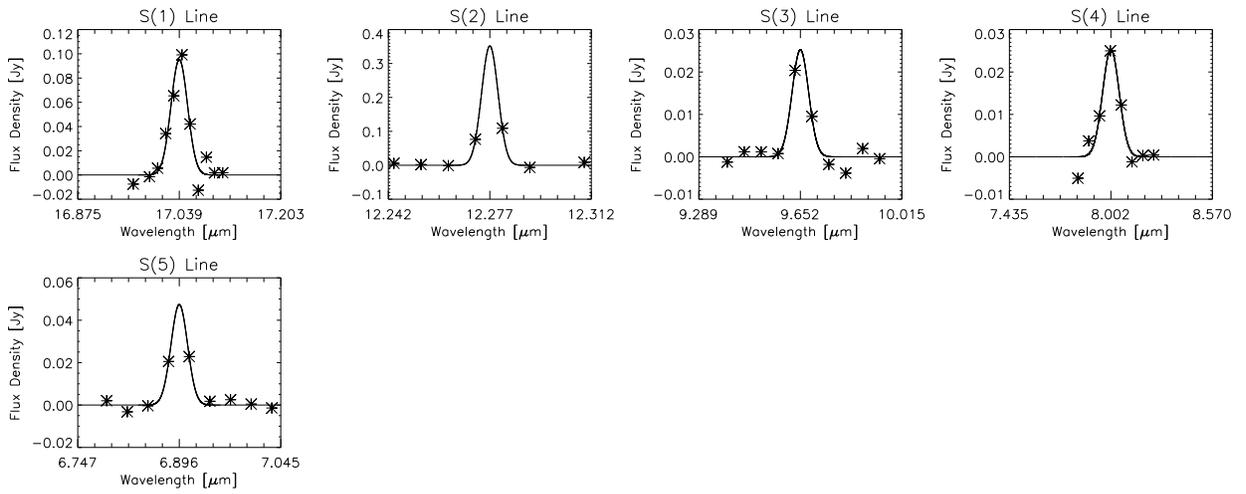}
   \caption{Same as Fig.~\ref{fig4} but for the spectrum close to RNO 1C. Here, the 12.28$\,\mu$m line was
   only detected in the high--resolution spectrum.}
 \label{fig5}%
\end{figure}

Keeping in mind the existence of a molecular outflow that is powered by IRAS 00338+6312 and directly oriented  
in the direction of RNO 1C and RNO 1B \citep[e.g.,][]{xu2006}, 
the detection of H$_2$ emission lines in both spectra hints towards shock induced emission related to the outflow. The spectra consequently bear information about 
the circumstellar material close to RNO 1B/1C and about the outflow coming from the IRAS source. Since we observe H$_2$ lines even in the 
10$\,\mu$m silicate absorption bands the outflow appears to lie in front of the dusty environment as otherwise the high extinction 
\citep[A$_V=9.2$ mag and A$_V\approx12.0$ mag for RNO 1B and RNO1C, respectively, ][]{staudeneckel1991} 
would have prevented a detection. The measurement of the relative strengths of multiple H$_2$ lines allows an analysis of the physical conditions of the shocked material. For this
we assume local thermal equilibrium (LTE) and optically thin line emission which is supported by the low Einstein coefficients of the involved quadrupole transitions
(Table~\ref{hydrogen_lines}). Following \cite{parmar1991} the column density of an upper energy level $N_u(J)$ is then given by
\begin{equation}\label{eq1}
N_u(J)=\frac{4\pi}{e^{-\tau}}\frac{I(J)}{A_{ul}\,\Delta E_{ul}}\;\rm{cm^{-2}}
\end{equation}
where $I(J)$ denotes the observed line intensity in erg s$^{-1}$ cm$^{-2}$ sr$^{-1}$, $\Delta E_{ul}$ is the energy 
difference of the two states involved in the transition, $A_{ul}$ is the
Einstein coefficient for the transition and $\tau$ is the optical depth at the observed wavelength. 
In Fig.~\ref{fig4} we show Gaussian fits 
to the observed emission lines in the spectrum close to RNO 1B and Fig.~\ref{fig5} presents corresponding fits to the lines 
detected close RNO 1C.
Before fitting the emission lines we subtracted the underlying continuum 
which was fitted with a second order polynomial.
The finally derived integrated line fluxes and column densities are shown in 
Table~\ref{hydrogen_lines}. 
We corrected the line fluxes for extinction effects using 
the results of \citet{mathis1990} and assuming $A_V=3.55$ mag. This value represents the extinction found by \citet{staudeneckel1991}
towards the nearby star RNO 1 and should be a better estimate for the line-of-sight extinction than the above mentioned extinction values towards RNO 1B and RNO 1C.
In any case the influence of the extinction on the derived shock temperatures and column densities (see below) is negligible. 
The 1--sigma errors in the line flux, the line intensity and the column density (Table~\ref{hydrogen_lines}) were derived from 
generating 500 spectra where we added a Gaussian noise to each measured flux point based on the initial individual 1--sigma uncertainty. 
From these spectra we computed the mean values and related errors of the listed parameters.
Especially at the short wavelength end it 
shows that not all lines were detected with a 3--sigma confidence level. However, since 
most lines were convincingly measured we decided to keep also the tentative 
detections in our analyses.

Since we assume LTE the rotational energy levels shall be populated following 
Boltzmann statistics with a unique temperature for several lines.
The involved temperatures $T_{\rm{rot}}$ of the shocked material can be derived from a so--called "rotational diagram". 
By using the results from Eq.~\ref{eq1}, plotting the logarithm of $N_u(J)/(g_s g_J)$ against $E_J/k$ 
(i.e. the formal temperature corresponding to the absolute upper energy level of the respective 
rotational transition) and fitting a straight line to the data one finds
that the slope of the straight line is proportional to $-1/T_{\rm{rot}}$. 
Here, $g_s$ denotes the spin degeneracy of each energy level (1 for even J, 3 for odd J) and $g_J=2J+1$ is the
rotational degeneracy. These numbers assume that the ortho--to--para ratio of the involved hydrogen is close to its LTE value of 3 at $T_{\rm{rot}}$.
A deviation from this LTE assumption (i.e., ortho--to--para $<$ 3) would result in a downwards displacement of the data points with an odd J-number (ortho--H$_2$) 
relative to the points with even J-numbers (para--H$_2$) and thus create a "zig--zag" pattern in the rotational plot \citep[e.g.,][]{neufeld2006,wilgenbus2000}.

In Fig.~\ref{fig6},~\ref{fig7}, and~\ref{fig8} we show  rotational plots for the measured lines near RNO 1B and RNO 1C. In all plots the errors for the data points denote the 1--sigma uncertainty in the measured column density.

The rotational diagram of RNO 1B in Fig.~\ref{fig6} shows a clear curvature in the data points. 
This departure from a single straight line 
is indicative of several temperature components in the shocked material. We thus fitted the data with two superimposed 
temperature regimes (hot and warm) that rather represent the minimum and maximum of the involved temperatures than any distinct intermediate value. 
The dash-dotted line fits the high energy regime and represents the hot component with a 
temperature of 2991$\pm$596 K. The 
dotted line corresponds to a temperature of 1071$\pm$121 K designating the warm component. 
The uncertainties in the temperature denote the 1--sigma confidence level of the
fits. As the derived temperature is extremely sensitive to the slope of the fitted straight
line the corresponding errors are rather large.

In addition to the temperatures one can also derive the total H$_2$ column density of the observed shock
within the given aperture. 
From the y-intersects of the fits 
we estimate the hot component to have $N^{\rm{hot}}_{\rm{1B}}$(H$_2$)$\approx 1.2\times 10^{17}\;\rm{cm^{-2}}$ and for the warm 
component we find $N^{\rm{warm}}_{\rm{1B}}$(H$_2$)$\approx 5.6\times 10^{18}\;\rm{cm^{-2}}$.

For the column densities derived from the spectrum close to RNO 1C we  
plotted two rotational diagrams shown in Fig.~\ref{fig7} and~\ref{fig8}. 
First, we fitted the data with a single temperature component 
of 1754$\pm$321 K. However, the shocked material is apparently not 
in LTE as the data points with odd and even J-numbers 
are difficult to fit simultaneously with a straight line or a 
curve. As mentioned above this displacement is indicative of a departure from the LTE ortho--to--para ratio of 3. 
Fig.~\ref{fig8} shows the same data if one assumes an ortho--to--para ratio of 1. Now, a two component fit, similar
to that in Fig.~\ref{fig6}, is possible. For the hot component we derive a temperature of 2339$\pm$468 K, while
the warm component is significantly colder with 466$\pm$264 K. The total H$_2$ column densities amount to 
$N^{\rm{hot}}_{\rm{1C}}$(H$_2$)$\approx 5.5\times 10^{17}\;\rm{cm^{-2}}$ and 
$N^{\rm{warm}}_{\rm{1C}}$(H$_2$)$\approx 9.7\times 10^{19}\;\rm{cm^{-2}}$, respectively.

Table~\ref{components} summarizes the temperatures and total column densities 
of the different components for both spectra. 

\section{Discussion}
The \emph{IRAC} images reveal for the first time the embedded MIR point source associated with IRAS 00338+6312.
However, already \citet{weintraubkastner1993} concluded from polarimetric observations that an additional source 
should be present close to the IRAS point source position.
Furthermore, many authors analyzing the bipolar outflow related to this region concluded that 
a deeply embedded object was most likely the driving source \citep{yang1995,xu2006,weintraubkastner1993} instead of one of the RNO objects \citep{mcmuldroch1995}. Previous ground based MIR observations \citep{polomski2005} 
were possibly not sensitive enough to detect IRAS 00338+6312.

The observed properties of IRAS 00338+6312 convincingly 
classify the object as a young protostar and support the idea that it is indeed driving the molecular 
outflow. From their CO measurements \citet{xu2006} derived a total mass for this bipolar
outflow of 1.4 M$_{\sun}$, indicating that IRAS 00338+6312 is an intermediate--mass protostar. The detection
of the IRAS source solves also the problem that up to now the SEDs of RNO 1B and RNO 1C 
were unusually steep longward of 25$\,\mu$m for FUOR objects \citep{weintraubkastner1993}. The images
presented here have sufficient sensitivity and spatial resolution to 
show that the newly revealed source associated with IRAS 00338+6312 is most probably 
responsible for this flux excess.

In addition to the IRAS object
\citet{weintraubkastner1993} also suspected another embedded source which we identify with RNO 1G.
Furthermore, the detection of RNO 1F \citep{weintraub1996}, which was initially assumed to be 
a density enhancement rather than an independent self-luminous source \citep{weintraubkastner1993},
is confirmed with our \emph{IRAC} images. Only RNO 1D, which was thought to be 
located between RNO 1B and RNO 1C \citep{staudeneckel1991}, does not
show up in our images. This source was, however, also 
not detected by \citet{weintraubkastner1993} and only tentatively seen in the data from \citet{weintraub1996}.

It is not clear whether all detected objects are physically confined to the  
RNO 1B/1C region. Especially the membership of the 
newly discovered objects RNO1 IRAC1, RNO1 IRAC2, and RNO1 IRAC3 needs to be confirmed.
However, the \emph{IRAC} observations suggest that RNO 1B and RNO 1C belong to a small cluster of 
(partly very) young objects. To our knowledge, although 
some FUORs are known or suspected to be in binary or multiple
systems \citep[e.g.,][]{reipurthaspin2004}, the existence of FUORs in a cluster-like environment  
is a so far unique finding. 
Only the FUOR candidate V1184 Tau (CB34) belongs also to a small cluster \citep{tigran2002}
but its classification as FUOR is far less certain than for the RNO objects \citep{yun1997}.
In any case the still deeply embedded protostars in the close vicinity of RNO 1B/1C put strong constraints 
on the age of the objects if one assumes coeval evolution.

The magnitudes we derive for the two FUOR objects at 3.6$\,\mu$m can be compared to previous 
ground based measurements 
at 3.8$\,\mu$m by \citet{kenyon1993} and \citet{polomski2005}. However, one certainly 
has to keep in mind that the \emph{IRAC} filter is not only different in terms of central 
wavelength but also the spectral width differs from that of the ground based instruments.
While \citet{kenyon1993} found $L_{3.8\,\mu\rm m}=6.36$\,mag and 
$L_{3.8\,\mu\rm m}=6.5\,$mag for RNO 1B and RNO 1C, respectively, \citet{polomski2005} observed
RNO 1C to be slightly brighter than RNO 1B ($L_{3.8\,\mu\rm m}=6.27\,$mag vs. $L_{3.8\,\mu\rm m}=6.42\,$mag).
Also in our measurements RNO 1C seems to be the brighter component. However, both objects appear
to be fainter in our observations compared to \citet{polomski2005}. 
Apart from the differences in the filter properties
intrinsic variations in the luminosity of the FUORs and/or changing local extinction effects can account for the
apparent variability in these objects. 

The detection of H$_2$ rotational line emission can in general be attributed to collisional excitation
from C- or J-type shocks \citep[e.g.,][]{drainemckee1993,moromartin2001}. C-type shocks are magnetic
and less violent towards molecules in comparison to the mostly hydrodynamic J-type shocks that
can dissociate H$_2$ molecules even at lower shock velocities.
From our rotational diagrams we derived shock temperatures that are similar to those found
in the outflow from Cepheus A \citep{froebrich2002}. These authors could fit a two component
C-shock model to their data and derived shock velocities between 25 and 30 km\,s$^{-1}$
for a cold and a hot component, respectively. By comparing our results directly
to theoretical shock models we find that the C-shock Model 1 from \citet{timmermann1998} can explain 
both the hot and warm component in the spectrum close to RNO 1B. Considering our derived temperatures
as an upper and lower limit for the shocked material, shock velocities 
between 15 and 30 km\,s$^{-1}$ are required. However, Model 1 of \citet{timmermann1998} predicts 
on ortho--to--para ratio of $\approx$2 at the low velocity/low temperature limit which is not
directly evident from our observations. More recent models by \citet{wilgenbus2000} predict 
lower shock temperatures in C-type shocks for the same shock velocities in comparison to
\citet{timmermann1998}. Following their models, our hot component in Fig.\ref{fig6} requires velocities 
exceeding 40 km\,s$^{-1}$. Slightly slower velocities ($\approx$35 km\,s$^{-1}$) are already needed to explain the
hot temperature component in the shock close to RNO 1C (Fig.~\ref{fig8}). The warm component in this diagram
represents shock velocities between 15 km\,s$^{-1}$ and 20 km\,s$^{-1}$. 
Although it is clear that some uncertainties between the observations and theory still remain, 
C-type shocks seem to provide a solid explanation for the observed shock temperatures. 

The measurement of an ortho--to--para ratio smaller than 3 in Fig.~\ref{fig8} implies that 
despite the high temperatures the gas has not yet reached equilibrium between the ortho and
para states. Thus, the observed ratio is the legacy of the temperature history of the gas
\citep[see, e.g.,][]{neufeld1998,neufeld2006} and transient heating by a recently passing shock wave
already caused higher temperatures. As the corresponding region on the sky lies closer
to the IRAS source than the region probed in Fig.~\ref{fig6}, the observations suggest that we 
see the signatures of at least two shock waves: The first shock wave is probed close to RNO 1B (Fig.~\ref{fig6})
where we observe several temperature components that have apparently reached their equilibrium state.
A more recent shock wave is seen close to RNO 1C (Fig.~\ref{fig8}) where the ratio of the line
column densities hints also towards high shock velocities and temperatures but the gas is not
yet in thermal equilibrium.

\section{Conclusions and Future Prospects}
Our conclusions can be summarized as follows:
\begin{itemize}
\item We detected and resolve for the first time the MIR point source associated with IRAS 00338+6312 
which appears to be 
an embedded intermediate--mass protostar driving the known
molecular outflow in the RNO 1B/1C region.
\item The detection of additional (partly previously unknown) point sources suggests
that the FUOR objects RNO 1B/1C belong to a young small stellar cluster.
To our knowledge, RNO 1B/1C are the only well--studied and confirmed FUORs that apparently belong to a 
cluster--like environment.
\item All but two objects were detected in all four \emph{IRAC} bands and their 
colors are consistent with Class 0/I-II objects. The two objects 
that were not detected at 3.6$\,\mu$m (including IRAS 00338+6312) are still 
very deeply embedded protostars.
\item Having apparently extremely young objects in the direct vicinity of the FUORs
confirms the suspected young age for this type of objects although their MIR colors are 
consistent with Class II objects.   
\item The two MIR spectra of the region bear clear signs of a dense icy and dusty
circumstellar environment as solid state features are seen in absorption.
\item The spectra show also H$_2$ emission lines from purely rotational transitions.
We presume that these lines arise from shocked material within the molecular outflow.
The derived shock temperatures and velocities are in agreement with C-type shock models.
\item The observations of the H$_2$ lines suggest that the outflow lies in front of RNO 1B/1C
as otherwise the high optical depth towards these objects would have prevented the detection.
\item While in one spectrum the gas probed by the H$_2$ line emission seems to be in LTE, the other
spectrum shows a deviation from the expected LTE ortho--to--para H$_2$ ratio. This indicates the
presence of at least two shock waves, the most recent one responsible for the non--LTE line ratios. 
\end{itemize}

The results presented here advantageously combine space-based MIR imaging and spectroscopy. While the
images give insights into the photometric properties of several young embedded objects and also 
FU Orionis type stars 
the spectra provide information on the dusty and icy circumstellar 
environment of the young cluster as well as on shocked gas within a molecular outflow. In this context the
existence of FUORs in a cluster--like environment needs to be pointed out. 
It is still debated whether all young low--mass stars undergo an FUOR phase or whether these objects form a special
sub--group of YSOs \citep[e.g.,][]{hartmannkenyon1996}. 
Our results suggests that the FUOR phenomenon occurs also in young (small) stellar clusters 
making it thus possibly more common than so far expected. In addition, the co--existence of FUORs and very young 
protostars in the same environment strengthens the idea that the FUOR--phase of YSOs is linked to the 
early stages of the star formation process. 

Possible future investigations of the RNO 1B/1C region should address whether the observed objects do 
indeed belong to the same region and whether all of them are young stars. Furthermore, a
detailed study of the shocked material including high--resolution narrow band imaging and spectroscopic mapping (e.g., with 
\emph{Spitzer IRS}) will provide deeper insights into its extension, physical conditions, 
and its relation to the molecular outflow. A complete analysis 
of the ice and dust features in the observed spectra is currently underway and will be
presented in an upcoming paper (Quanz et al. in preparation).







\acknowledgments
S. P. Quanz kindly acknowledges support from the German
\emph{Friedrich-Ebert-Stiftung}. We thank H. Beuther, R. Mundt, A. Carmona and S. Birkmann for
useful discussions and conversations. An anonymous referee provided valuable criticism that helped improve
the paper. This work is based on observations made with the Spitzer Space Telescope, 
which is operated by the Jet Propulsion Laboratory, California Institute of Technology under a contract with NASA. This research has made use of the SIMBAD database,
operated at CDS, Strasbourg, France. 



{\it Facilities:} \facility{Spitzer}
\appendix
\section{The non-detection of the 28.22$\,\mu$m emission line}
\begin{figure}
 \vspace{1.5cm}
\centering
  \epsscale{0.8}
 \plotone{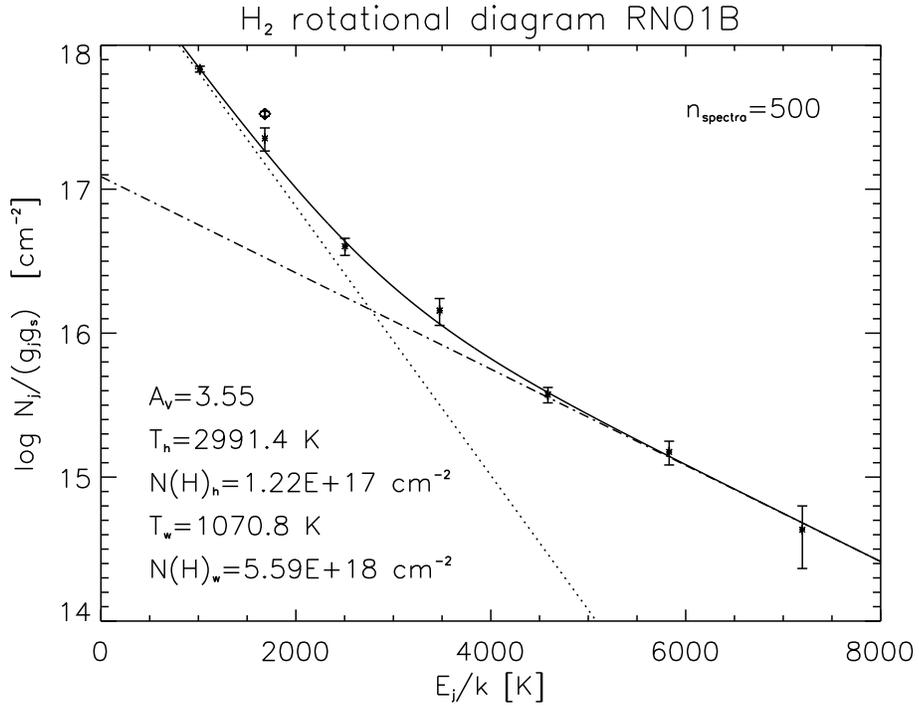}
   \caption{Rotational diagram of the H$_2$ lines shown in Fig.~\ref{fig4}. 
   The data are well constrained by the superposition of a hot and a warm 
  component providing an upper and a lower limit for the temperature range of the shocked gas.
  The diamond at $\approx$1682 K corresponds to the 12.28$\,\mu$m line measured in the high--resolution module.
  The observed line flux, and hence the derived column density, 
  is slightly higher than that in the low--resolution spectrum. For the fit, however, the
  low--resolution data point was used as all other higher energy data points were also obtained with 
  the low--resolution spectrograph and uncertainties related to the different apertures and their orientations  
  for the high-- and low--resolution modules are thus minimized.} 
 \label{fig6}%
\end{figure}

\begin{figure}
 \vspace{1.5cm}
\centering
  \epsscale{0.8}
 \plotone{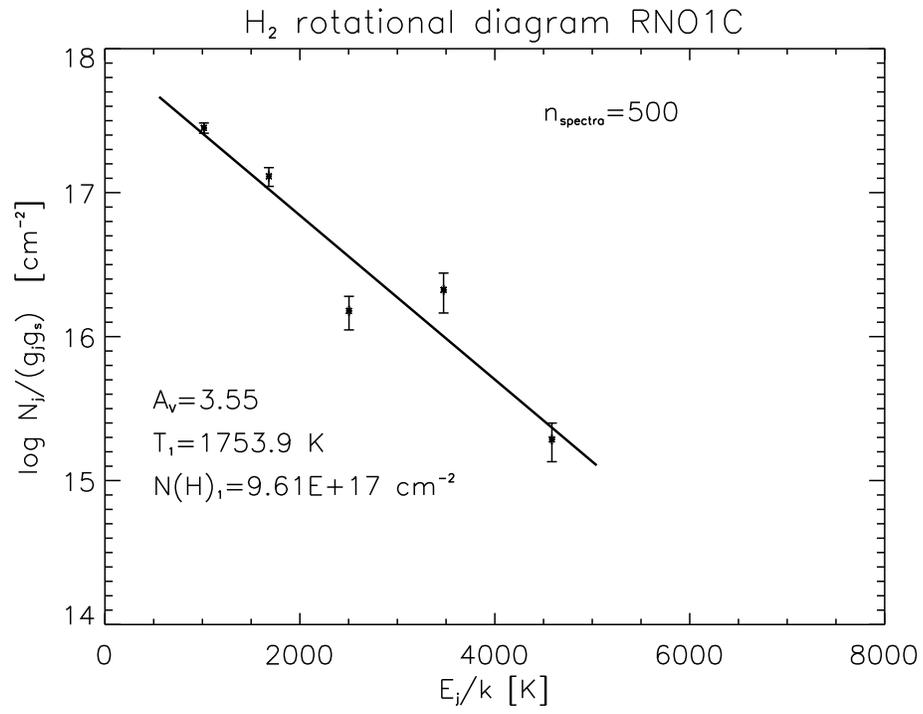}
   \caption{Rotational diagram of the H$_2$ lines shown in Fig.~\ref{fig4}. 
   The apparent zig-zag pattern is indicative of an ortho--to--para ratio smaller than 3.} 
 \label{fig7}%
\end{figure}

\begin{figure}
 \vspace{1.5cm}
\centering
  \epsscale{0.8}
 \plotone{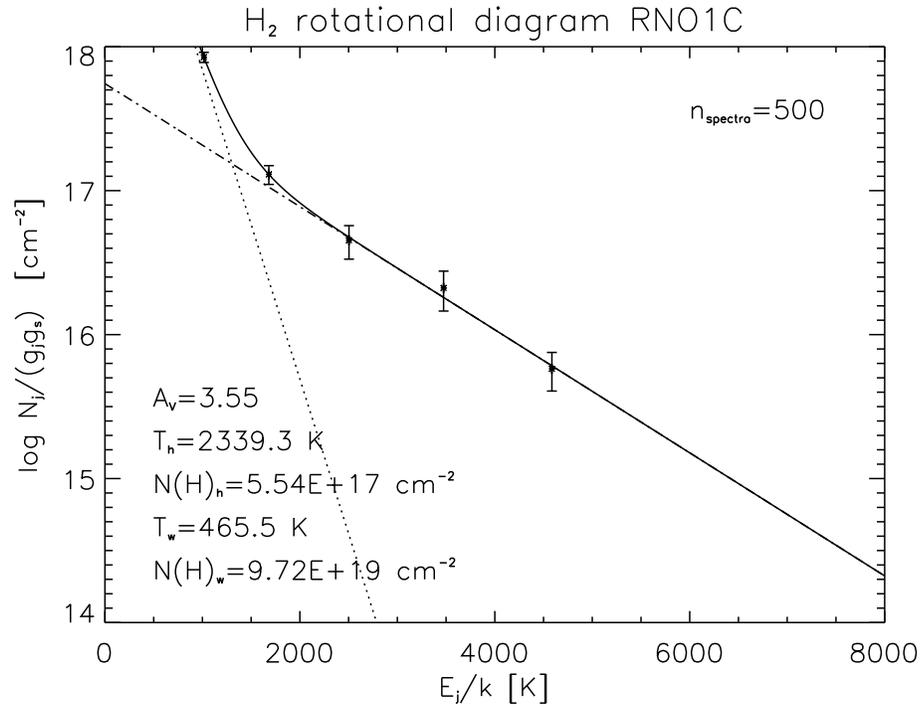}
   \caption{Same as Fig.~\ref{fig7} but now for an assumed ortho--to--para ratio of 1.
   A two component fit similar to that in Fig.~\ref{fig6} is now possible.} 
 \label{fig8}%
\end{figure}

\begin{figure}
 \vspace{1.5cm}
\centering
  \epsscale{1.}
 \plotone{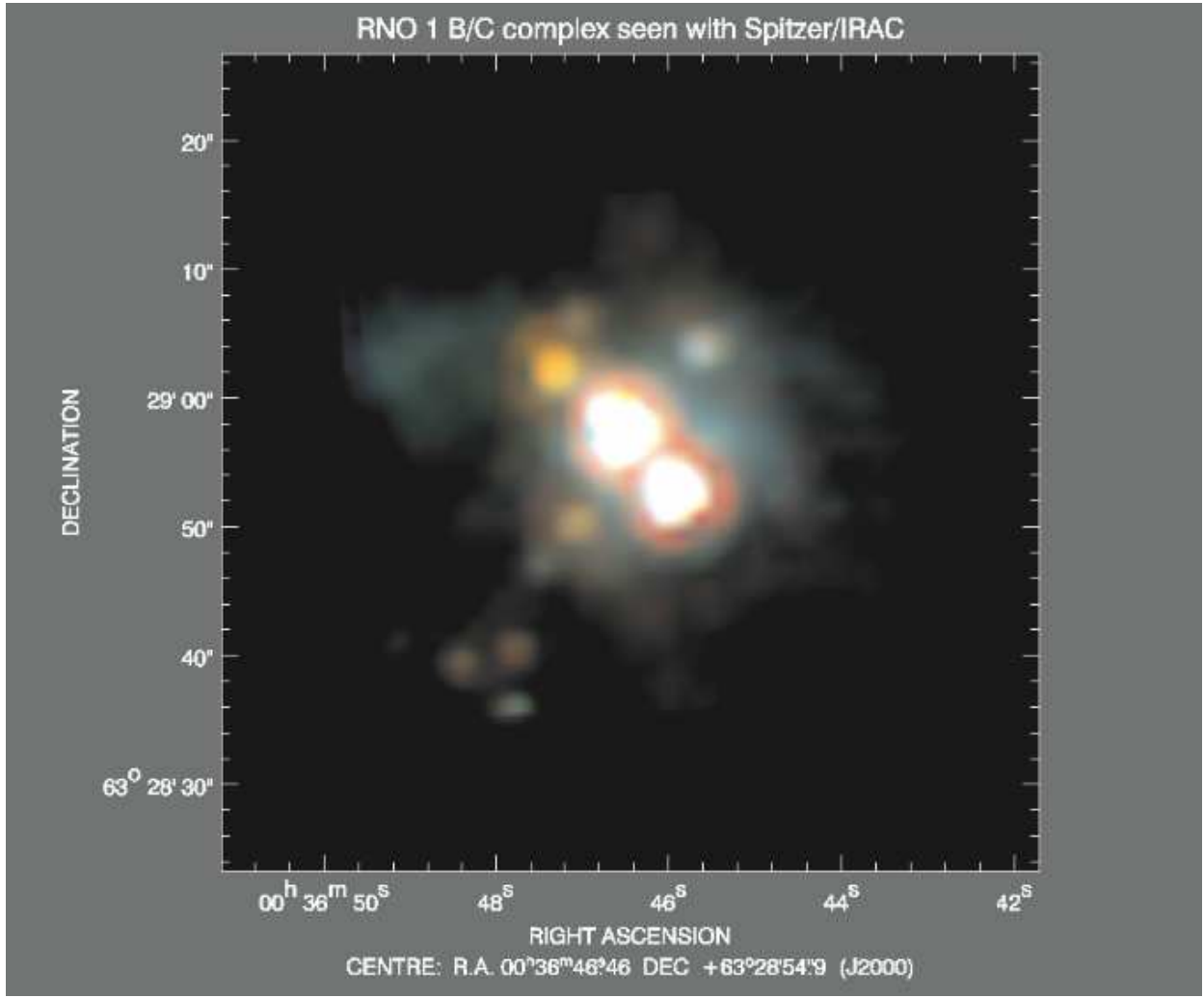}
   \caption{Color composite of the RNO 1B/1C region based on the three \emph{IRAC} bands at 3.6, 4.5, and
   8.0$\,\mu$m. The image is centered on RNO 1C. We applied a logarithmic color stretch to
   enhance the fainter sources. The extended emission east of the IRAS
    source which is predominantly seen in the green 4.5$\,\mu$m filter is possibly related to H$_2$ emission 
    within the bipolar outflow of this object. } 
 \label{fig9}%
\end{figure}

In the following we analyze whether the non-detection of the 28.22$\,\mu$m in either spectrum 
is in agreement with the models. The two component models in Fig.~\ref{fig6} and ~\ref{fig8} 
can be used to predict the column densities of the S(0) emission line. One expects 
$$\rm{log}\frac{N_{\rm{S(0)}}}{g_jg_s}[\rm{cm}^{-2}]\approx 18.3\,\rm{cm}^{-2}$$
$$\rm{log}\frac{N_{\rm{S(0)}}}{g_jg_s}[\rm{cm}^{-2}]\approx 18.8\,\rm{cm}^{-2}$$
for the spectrum close to RNO 1B and RNO 1C, respectively.
With $g_s$=1 and $g_j$=5 and applying Eq.~(\ref{eq1}) the following line intensities are derived: 
$$\rm I(2)_{\rm{1B}}\approx 1.64\cdot10^{-6}\,\rm{erg\,s^{-1}\,cm^{-2}\,sr^{-1}}$$
$$\rm I(2)_{\rm{1C}}\approx 5.20\cdot10^{-6}\,\rm{erg\,s^{-1}\,cm^{-2}\,sr^{-1}}$$
The $A$-coefficient for the transition is $2.94\cdot10^{-11}$s$^{-1}$ \citep{wolniewicz1998}.

Taking into account the aperture size for the long wavelength high--resolution module of the spectrograph
of 247.53 arcsec$^2$ ($\approx 5.82\cdot10^{-9}\,\rm{sr}$) the predicted integrated line fluxes amount to:
$$\rm F_{\rm{1B}}\approx 9.57\cdot10^{-22}\,\rm W\,\rm{cm}^{-2}$$
$$\rm F_{\rm{1C}}\approx 3.02\cdot10^{-21}\,\rm W\,\rm{cm}^{-2}$$

To estimate the peak of the 28.22$\,\mu$m line in Jansky we assume that the spectral resolution in the 
high--resolution modules is constant and that the FWHM of the S(0) line can be extrapolated from 
the FWHM of the S(1) line at 17.03$\,\mu$m. Since we fit a Gaussian profile to the emission
lines and we want to derive the peak flux of this Gaussian, 
we also have to take into account the relation between the FWHM we measure for the line 
and the $\sigma$ of the profile. This relation is FWHW=$\sqrt{8\,\rm{ln}2}\cdot\sigma$. 
With the expected FWHM of
$$\Delta\nu_{\rm{1B}}\approx 2.84\cdot10^{10}\,\rm{Hz}$$
$$\Delta\nu_{\rm{1C}}\approx 2.12\cdot10^{10}\,\rm{Hz}$$
we find thus corresponding peak fluxes of:
$$\rm F^{\rm{peak}}_{\rm{1B}}\approx 0.034\,\rm{Jy}$$
$$\rm F^{\rm{peak}}_{\rm{1B}}\approx 0.152\,\rm{Jy}$$

These values have to be compared to the measured uncertainties in the spectra. The mean 1--$\sigma$
level in the spectral range between 28.0 and 28.4$\,\mu$m is, however, 
$$\bar\sigma_{\rm{1B}}=0.055$$
$$\bar\sigma_{\rm{1C}}=0.082$$
so that even in the more favorable case of RNO 1C the emission line is not expected to be detected
with a confidence level greater than $\approx$1.8$\,\sigma$.

\clearpage

\begin{deluxetable}{llllcccc}
\rotate
\tablecaption{Journal of observations. \label{journal}}           
\tablewidth{0pt}
\tablehead{
\colhead{Instrument} & \colhead{RA (2000)\tablenotemark{a}} & \colhead{DEC (2000)\tablenotemark{a}} & \colhead{AOR Key} & \colhead{Filter or Module} & 
\colhead{Ramp duration /}  &  \colhead{Frame Time /} &  \colhead{Date}\\
 & & & & & \colhead{\# of Cycles} & \colhead{\# of Frames /} & \\
  & & & &&  & \colhead{Dither Positions} & 
}
\startdata
Spitzer IRAC&   00:36:45.8 &    +63:28:56   & 5027072 & 3.6, 4.5, 5.8, 8.0$\,\mu$m &  - & 0.4 / 1 / 4 & 2003-12-20\\
Spitzer IRS&   00:36:46.34\tablenotemark{b}    &  +63:28:53.76\tablenotemark{b}  &6586624 & Low Resolution & 6 sec / 3 &  - &  2004-01-07\\
		&    00:36:46.89\tablenotemark{c}   &  +63:28:58.44\tablenotemark{c}      & & & & \\
Spitzer IRS&     00:36:46.34\tablenotemark{b}    &  +63:28:53.76\tablenotemark{b}      & 6586624 & High Resolution & 6 sec / 5 &  - &  2004-01-07\\
		&    00:36:46.89\tablenotemark{c}   &  +63:28:58.44\tablenotemark{c}      & & & & 
\enddata
\tablenotetext{a}{Average slit position of low--resolution spectrograph.}
\tablenotetext{b}{close to RNO 1B}
\tablenotetext{c}{close to RNO 1C}

\end{deluxetable}

\clearpage

\begin{deluxetable}{clcccccc}
\rotate
\tablecaption{Apparent brightness at different wavelengths for objects that were detected in at least three
IRAC bands. 
The coordinates are measured in the 3.6$\,\mu$m image. 
Only the positions of the IRAS source and RNO 1G were measured in 
the 4.5$\,\mu$m exposure.\label{irac_fluxes}}           
\tablewidth{0pt}
\tablehead{
\colhead{No.} & \colhead{Object name} & \colhead{RA (J2000)} & \colhead{DEC (J2000)} & \colhead{3.6$\,\mu$m} & \colhead{4.5$\,\mu$m} & 
\colhead{5.8$\,\mu$m} & \colhead{8.0$\,\mu$m}\\
& & & & \colhead{[mag]} & \colhead{[mag]} & \colhead{[mag]} &\colhead{[mag]}}
\startdata
1 & RNO 1B & 00:36:46.05 & +63:28:53.29 & 7.16$\pm$0.13 & 6.67$\pm$0.06 & 5.76$\pm$0.12 & 5.01$\pm$0.09\\
2 & RNO 1C & 00:36:46.65 & +63:28:57.90 & 6.56$\pm$0.01 & 6.04$\pm$0.01 & 5.58$\pm$0.01 & 4.61$\pm$0.02\\
3 & RNO 1F & 00:36:45.74 & +63:29:04.09 & 10.28$\pm$0.09 & 9.46$\pm$0.08 & 8.58$\pm$0.05 & 8.20$\pm$0.03\\
4 & RNO 1G\tablenotemark{a} & 00:36:47.14 & +63:28:49.95 & - & 10.33$\pm$0.06 & 8.74$\pm$0.04 & 8.05$\pm$0.07\\
5 & IRAS 00338+6312 & 00:36:47.34 & +63:29:01.61  & - & 9.05$\pm$0.07 & 7.19$\pm$0.05 & 6.72$\pm$0.03\\
6 &RNO1 IRAC1 & 00:36:48.44 & +63:28:39.98 & 13.72$\pm$0.15 & 11.82$\pm$0.14 & 10.93$\pm$0.20 & 9.65$\pm$0.04\\
7 & RNO1 IRAC2 & 00:36:47.90 & +63:28:36.30 & 12.00$\pm$0.08 & 10.86$\pm$0.10 & 10.86$\pm$0.16 & 9.82$\pm$0.14 \\
8 & RNO1 IRAC3 & 00:36:47.85 & +63:28:41.23 & 13.07$\pm$0.10 & 11.78$\pm$0.08 & 10.74$\pm$ 0.10 & 9.74$\pm$0.05 \\
\enddata
\tablenotetext{a}{Embedded YSO from \citet{weintraubkastner1993}}

\end{deluxetable}

\clearpage
\thispagestyle{empty}
\begin{deluxetable}{lccccccc}
\rotate
\tablecaption{Properties of the observed Hydrogen emission 
lines close to RNO 1B (upper part) and RNO 1C (lower part). 
The numbers in parentheses denote powers of ten. The 12.28$\,\mu$m emission line close to RNO 1B was
measured in the low-- and high--resolution part of the spectrograph. See text for discussion about the apparent flux 
difference between the two measurements. \label{hydrogen_lines}}           
\tablewidth{0pt}
\tablehead{
\colhead{Transition} &  \colhead{Wavelength} & \colhead{Energy\tablenotemark{a}} & \colhead{$A$-Coefficient\tablenotemark{b}} &  
\colhead{Beam Size\tablenotemark{c}} & \colhead{Line Flux\tablenotemark{d}} & \colhead{Line Intensity\tablenotemark{d}} & \colhead{Column Density\tablenotemark{d,e}}\\
\colhead{} & \colhead{[$\,\mu$m]} & \colhead{E$_J$/k [K]} & \colhead{[s$^{-1}$]} & \colhead{[arcsec$^2$]} & \colhead{[W cm$^{-2}$]} 
& \colhead{[erg cm$^{-2}$ s$^{-1}$ sr$^{-1}$]} & \colhead{[cm$^{-2}$]}
}
\startdata
\multicolumn{8}{c}{H$_2$ lines close to RNO 1B} \\\hline
S(1) J=3-1 & 17.0348 & 1015.08 & 4.76(-10) & 53.11   & 7.91(-21)$\pm$3.8(-22)   & 6.34(-5)$\pm$3.0(-6) &  1.44(19)$\pm$6.8(17) \\
S(2) J=4-2 & 12.2786 & 1681.63 & 2.75(-9)  & 53.11   & 1.33(-20)$\pm$5.2(-22)   & 1.06(-4)$\pm$4.1(-6) & 3.00(18)$\pm$1.2(17) \\
           &         & 	       & 	   & 39.96   & 6.75(-21)$\pm$1.2(-21)   & 7.18(-5)$\pm$1.3(-5) & 2.03(18)$\pm$3.7(17) \\
S(3) J=5-3 & 9.6649 & 2503.73 & 9.83(-9)  & 39.96   & 2.00(-20)$\pm$2.7(-21)     & 2.13(-4)$\pm$2.9(-5) &  1.33(18)$\pm$1.8(17) \\
S(4) J=6-4 & 8.0251 & 3474.48 & 2.64(-8)  & 39.96   & 9.12(-21)$\pm$1.9(-21)     & 9.71(-5)$\pm$2.1(-5) & 1.87(17)$\pm$4.0(16)\\
S(5) J=7-5 & 6.9095 & 4585.94 & 5.88(-8)  & 39.96   & 2.13(-20)$\pm$2.6(-21)     & 2.26(-4)$\pm$2.8(-5) & 1.68(17)$\pm$2.1(16) \\
S(6) J=8-6 & 6.1086 & 5829.66 & 1.14(-7)  & 39.96   & 7.04(-21)$\pm$1.3(-21)     & 7.50(-5)$\pm$1.4(-5) & 2.54(16)$\pm$4.8(15)\\
S(7) J=9-7 & 5.5112 & 7196.20 & 2.00(-7)  & 39.96   & 1.32(-20)$\pm$6.1(-21)     & 1.41(-4)$\pm$6.5(-5) & 2.46(16)$\pm$1.1(16) \\\\
\multicolumn{8}{c}{H$_2$ lines close to RNO 1C} \\\hline
S(1) J=3-1 & 17.0348 & 1015.08 & 4.76(-10) & 53.11 & 3.26(-21)$\pm$2.7(-22)  & 2.61(-5)$\pm$2.1(-6) &  5.92(18)$\pm$4.8(17) \\
S(2) J=4-2 & 12.2786 & 1681.63 & 2.75(-9)  & 53.11   & 5.17(-21)$\pm$7.7(-22)   & 4.14(-5)$\pm$6.2(-6) & 1.17(18)$\pm$1.7(17) \\
S(3) J=5-3 & 9.6649 & 2503.73 & 9.83(-9)  & 39.96   & 7.52(-21)$\pm$2.0(-21)     & 8.00(-5)$\pm$2.1(-5) &  4.98(17)$\pm$1.3(17) \\
S(4) J=6-4 & 8.0251 & 3474.48 & 2.64(-8)  & 39.96   & 1.34(-20)$\pm$4.1(-21)     & 1.43(-4)$\pm$4.4(-5) & 2.74(17)$\pm$8.5(16)\\
S(5) J=7-5 & 6.9095 & 4585.94 & 5.88(-8)  & 39.96   & 1.10(-20)$\pm$3.3(-21)     & 1.17(-4)$\pm$3.5(-5) & 8.69(16)$\pm$2.6(16) \\
\enddata
\tablenotetext{a}{Energy level of the upper state following \citet{jennings1987}}
\tablenotetext{b}{\citet{wolniewicz1998}}
\tablenotetext{c}{Corresponds to the instrument aperture size in case 
of the short high--resolution module (53.11 arcsec$^2$, see \emph{Spitzer} Handbook). For all other transitions (i.e., 
short low--resolution module) the number denotes the extracted beam size during data reduction process (39.96 arcsec$^2$).}
\tablenotetext{d}{Corrected for extinction using $A_V=3.55$ mag.} 
\tablenotetext{e}{Upper energy state} 
\end{deluxetable}

\clearpage

\begin{deluxetable}{lcc}
\rotate
\tablecaption{Upper and lower limits on the temperatures, and the total H$_2$ column 
densities of the observed shocked components. 
For the data derived close to RNO 1C the results from the two component fit and the single component fit with 
different ortho--to--para ratios are given.\label{components}}           
\tablewidth{0pt}
\tablehead{
\colhead{Component} & \colhead{T [K]} & \colhead{log$_{10}$ N(H$_2$) [cm$^{-2}$]}

}
\startdata
\multicolumn{3}{c}{RNO 1B} \\\hline
Hot\tablenotemark{a} & 2991$\pm$596 & 17.09$\pm$0.36\\
Warm\tablenotemark{a}& 1071$\pm$121 & 18.75$\pm$0.09\\\\
\multicolumn{3}{c}{RNO 1C} \\\hline
Hot\tablenotemark{b} & 2339$\pm$468 & 17.74$\pm$0.31\\
Warm\tablenotemark{b} & 466$\pm$264 & 19.99$\pm$1.17\\\\\hline
Single\tablenotemark{a} & 1754$\pm$321 & 17.98$\pm$0.31\\\\
\enddata
\tablenotetext{a}{ortho--to--para ratio of 3} 
\tablenotetext{b}{ortho--to--para ratio of 1} 
\end{deluxetable}

\clearpage

\clearpage

\clearpage

\clearpage

\clearpage

\clearpage

\clearpage

\clearpage

\clearpage

\clearpage

\end{document}